\documentclass[letterpaper]{IEEEtran}
\IEEEoverridecommandlockouts
\usepackage{algorithm}
\usepackage{cite}
\usepackage{amsmath,amssymb,amsfonts}
\usepackage{algorithmic}
\usepackage{graphicx}
\usepackage{subfigure}
\usepackage{textcomp}
\usepackage{xcolor}
\usepackage{gensymb}
\usepackage{multirow}
\usepackage{booktabs}  % ºÏ²¢µ¥Ô
\usepackage{threeparttable}
\usepackage{epstopdf}
\usepackage{amssymb}
\usepackage{makecell}
\usepackage{upgreek}
\usepackage{bm}
\usepackage{multirow}
\usepackage{makecell}
\usepackage{url}
\usepackage{dblfloatfix}
\makeatletter

\renewcommand{\maketag@@@}[1]{\hbox{\m@th\normalsize\normalfont#1}}%
\makeatother

\def\BibTeX{{\rm B\kern-.05em{\sc i\kern-.025em b}\kern-.08em
    T\kern-.1667em\lower.7ex\hbox{E}\kern-.125emX}}
\begin{document}
\title{Environment Reconstruction with Multi-targets Reflectors-merged Sensing Method Based on THz Single-sided Channel Characteristics
\thanks{Manuscript received xxx; revised xxx; accepted xxx. Date of publication xxx; data of current version xxx. This work was supported in part by the National Natural Science Foundation of China (Nos. 62201086, 62101069, and 92167202), the National Science Fund for Distinguished Young Scholars (No. 61925102), and BUPT-CMCC Joint Innovation Center. $\textit{(Corresponding \ auther: \ Jianhua \ Zhang.)}$\\
Zhaowei Chang, Pan Tang, and Jianhua Zhang are with the State Key Lab of Networking and Switching Technology, Beijing University of Posts and Telecommunications, Beijing 100876, China (e-mail: \{changzw12345, tangpan27, jhzhang\}@bupt.edu.cn).\\
Hao Jiang is with the School of Artificial Intelligence/School of Future Technology, Nanjing University of Information Science and Technology, Nanjing 210044, China; and also with the National Mobile Communications Research Laboratory, Southeast University, Nanjing 210096, China (e-mail: jianghao@nuist.edu.cn).\\
Guangyi Liu is with China Mobile Research Institute, Beijing 100053, China (e-mail: liuguangyi@chinamobile.com).
}
}

\author{\IEEEauthorblockN{Zhaowei Chang, Pan Tang, Jianhua Zhang, Hao Jiang, Guangyi Liu
}\\
%\IEEEauthorblockA{$^1$State Key Lab of Networking and Switching Technology, Beijing University of Posts and Telecommunications, Beijing 100876, China}\\
%\IEEEauthorblockA{$^2$China Mobile Research Institute,
%Beijing, 100053, China}\\
%\IEEEauthorblockA{Email: \{changzw12345, jhzhang, 
% tangpan27, tianlbupt, yadongyang, linjx\}@bupt.edu.cn, liuguangyi@chinamobile.com}
}
\maketitle

\begin{abstract}
Terahertz (THz) integrated sensing and communication (ISAC) holds the potential to achieve high data rates and high-resolution sensing. Reconstructing the propagation environment is a vital step for THz ISAC, as it enhances the predictability of the communication channel to reduce communication overhead. In this letter, we propose an environment reconstruction methodology (ERM) merging reflectors of multi-targets based on THz single-sided channel small-scale characteristics. In this method, the inclination and position of tiny reflection faces of one single multi-path (MPC) are initially detected by double-triangle equations based on Snell’s law and geometry properties. Then, those reflection faces of multi-target MPCs, which are filtrated as available and one-order reflection MPCs, are globally merged to accurately reconstruct the entire propagation environment. The ERM is capable of operating with only small-scale parameters of receiving MPC. Subsequently, we validate our ERM through two experiments: bi-static ray-tracing simulations in an L-shaped room and channel measurements in an urban macrocellular (UMa) scenario in THz bands. The validation results demonstrate a small deviation of 0.03 m between the sensing outcomes and the predefined reflectors in the ray-tracing simulation and a small sensing root-mean-square error of 1.28 m and 0.45 m in line-of-sight and non-line-of-sight cases respectively based on channel measurements. Overall, this work is valuable for designing THz communication systems and facilitating the application of THz ISAC communication techniques.
\end{abstract}


\section{INTRODUCTION}\label{1}
\IEEEPARstart{I}{ntegrating} sensing into communication is considered to be enhanced capabilities envisioned for sixth generation (6G) communication system \cite{sci_china}. To meet the essential requirements, integrated sensing and communication (ISAC) enables base stations (BSs) or terminals to simultaneously communicate and sense the surrounding environment \cite{isac_back}, \cite{yam}. There will be a high demand for high transmission rates for communication and sensing. THz bands with theirs wide bandwidth of 0.1-10 THz could provide high-resolution in the delay domain and high-speed data rates ranging from tens of Gbps to several Tbps for ISAC \cite{THz_1}, \cite{THz_2}. Additionally, as the frequency increases to the THz range, the shorter wavelength closely interacts with particles in the environment, leading to less pronounced diffraction and prevailing reflection in the THz bands \cite{fitee_6g}. This simplification in complexity and improvement in accuracy of sensing by specifically identifying the reflection path, thereby making the combination of both THz and ISAC an available and ideal solution for achieving ubiquitous sensing in 6G. To facilitate the sensing-assisted function of THz ISAC, it is essential to acquire environmental information. This information can enhance communication quality by making the propagation channel more predictable. It can be utilized for optimization techniques or system configuration adjustments at both the BS and user equipment (UE), for example, updating the direction of spatial filtering in beamforming to mitigate significant occlusion loss or prevent link failure. This necessitates the initial step, environment reconstruction (ER), to provide the actual environment information in THz ISAC. Thus, it is imperative to investigate the ER methodology (ERM) with support from channel characteristics in a realistic THz communication system. 

Recently, a multitude of investigations have been conducted for ER. The majority of existing research is limited to investigating a single reflector in the environment. For instance, mono-static experiments at 28 $\rm{GHz}$ \cite{isac_28ghz} and 330 $\rm{GHz}$ \cite{isac_human} have been conducted to explore the human torso influence and imaging, respectively. Besides, the reconfigurable intelligent surface (RIS) positioning is conducted at vehicle-to-vehicle (V2V) wireless systems\cite{ris_posi}, and the radar cross-section (RCS) of one wall in the environment is studied in mono-static at the 215-225 $\rm{GHz}$ frequency bands \cite{isac_215ghz} for ER. There is also some research on reconstructing the whole environment under mono-static radar cases. An indoor environment is generated by an unmanned aerial vehicle (UAV) \cite{ISAC_v2v}, mono-static ER are conducted in an open room and a conference room measurements at 140 $\rm{GHz}$ in \cite{isac_hanchong} and \cite{huawei}, respectively. In summary, current research on ER is predominantly centered around a single reflector target. While most research focuses on mono-static ER of simple targets, there is a limited emphasis on reconstructing the entire propagation environments in communication systems, which are commonly bi-static. 

To address these challenges, in this letter, we propose an ERM to sense and merge multi-targets reflectors of all multi-paths (MPCs) based on single-sided channel small-scale parameters. The ray-tracing simulation and an extensive measurement campaign in the typical UMa scenario in THz bands are conducted to validate our methodology. Subsequently, we investigate the MPCs mapping relations to realistic environmental layouts in the spatial domain by delay-angular-power-spectrum (DAPS). The method verification demonstrates a small root-mean-square error (RMSE), indicating high accuracy. The contributions of this letter can be summarized as follows:
\begin{itemize}
\item A ERM is proposed for reconstructing the propagation environment using small-scale parameters of the receiving radio in the context of the THz ISAC application, which aims at achieving the environmental positions to enhance communication quality. 
\item To validate the fundamental geometric derivation of ERM, a ray-tracing simulation is performed in an L-shaped room. The minimal deviation of 0.02 between the simulated reflection points (RPs) and reconstructing one demonstrates the high accuracy of our method.
\item An extensive channel measurement campaign is conducted in the UMa scenario in THz bands to validate the feasibility of the proposed method in an actual communication system. The reconstruction results showed small RMSEs, indicating that the ERM performs well.
\end{itemize}
\section{The Multi-target Reflectors-merged ERM}\label{4}
The radio information of UE in THz bistatic communication not only just encompasses the THz channel characteristics, but also possesses the capability to sense the surrounding environment. This is attributed to its straightforward propagation mechanism primarily based on reflection or scattering. In this subsection, we propose a reconstruction deduction of multi-reflectors for each UE.
\begin{figure}[!t]
\centerline{\includegraphics[width=6cm, trim=2.1cm 2.8cm 4.95cm 0.05cm, clip]{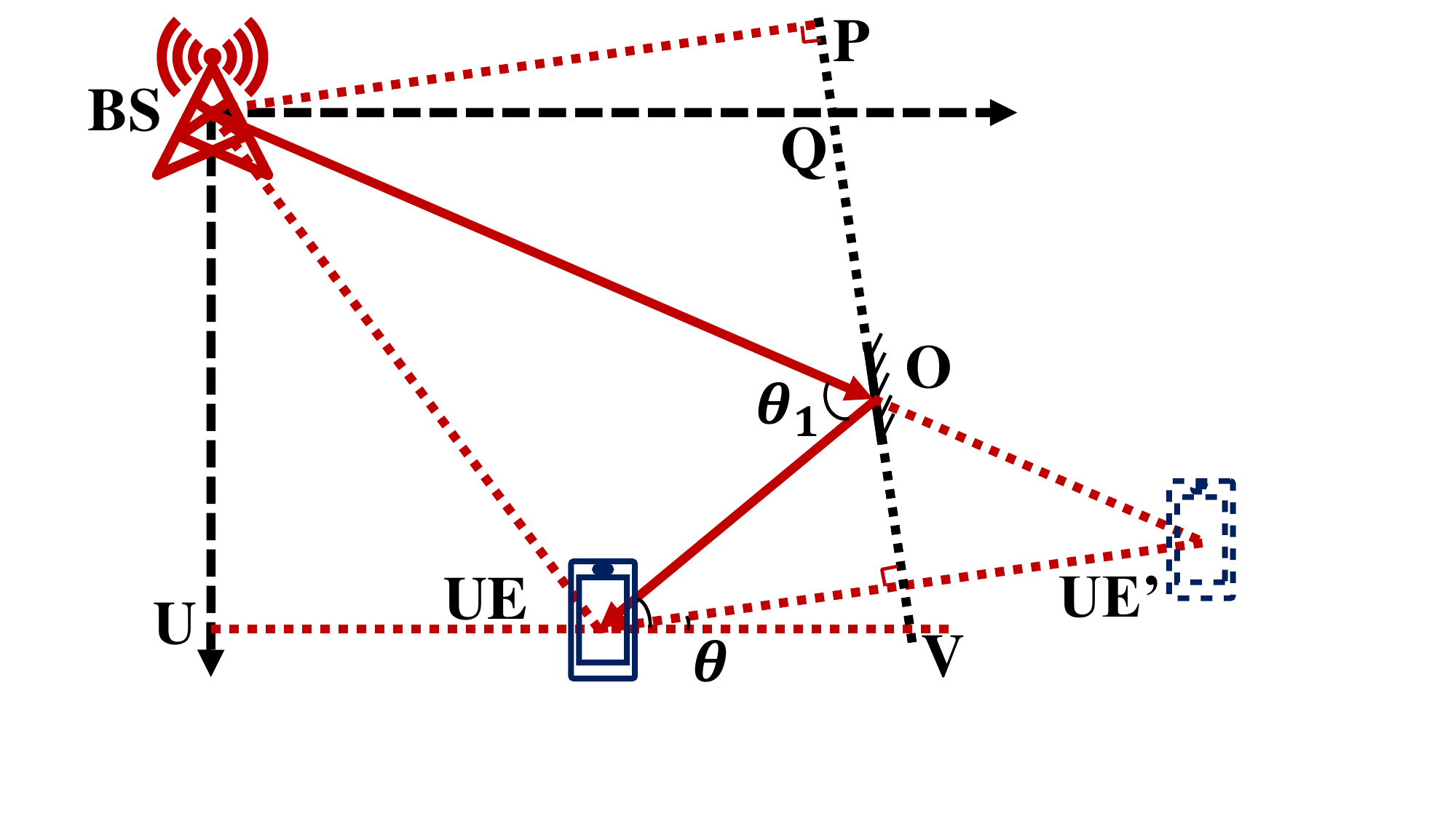}}
\caption{The layout of the propagation path and the reflection scenario.}\label{sensing layout}
\end{figure}

To explicitly elucidate the geometric relationship among the RPs, the propagation path, and the reflection surface, Fig. \ref{sensing layout} shows a specific example of the reflection scenario for a single MPC. The derivation process for other MPCs and reflectors at different positions follows a similar approach. The RP is deduced by the mirror symmetry of the UE'. The red dot lines represent auxiliary lines, while the black dot lines and dash lines denote extension lines of the reflector and coordinate axis, respectively. Any reflection surfaces on the building facade are represented by solid black lines. The distances of BS-U $d_\text{BS-U}$ and U-UE $d_\text{U-UE}$ can be obtained through measurement and calculation. Besides, $\angle$O-UE-V and the propagation distance BS-O-UE $d_\text{BS-O-UE}$ are the azimuth angle of arrival (AoA) and the delay$\times3\times10^8$ m of the MPC, respectively. Two equations based on Snell's law and geometry can be derived from this analysis. One is the basic theorem of Euclidean plane geometry in $\bigtriangleup$BS-UE-O. The equation is written as:
\begin{equation}\label{sensing eq1}
\begin{split}
d_\text{BS-O}^2+d_\text{UE-O}^2-2d_\text{BS-O}d_\text{UE-O}\cos{\theta_\text{1}}=d_\text{BS-UE}^2,
\end{split}
\end{equation}
\noindent where $d_\text{BS-UE}=\sqrt{d_\text{BS-U}^2+d_\text{U-UE}^2}$, $d_\text{UE-O}$ is the unknown parameters to be solved, the $d_\text{BS-O}=d_\text{BS-O-UE}-d_\text{UE-O}$, and $\theta_\text{1}$ represents the radian of $\angle$BS-O-UE which can be expressed as: 
\begin{equation}\label{sensing theta1}
\begin{split}
\angle\text{BS-O-UE}^\circ&=180^\circ-\angle\text{UE-O-V}^\circ-\angle\text{BS-O-P}^\circ\\
&=180^\circ-2\times\angle\text{UE-O-V}^\circ\\
&=180^\circ-2\times(90^\circ - \angle\text{O-UE-V}^\circ+\theta^\circ)\\
&=2\angle\text{O-UE-V}^\circ-2\theta^\circ,
\end{split}
\end{equation}
\noindent where $\angle\text{UE'-UE-V}$ $\theta$ is also the unknown parameters to be solved and the superscript $\circ$ denotes the unit of degree. The other equality condition is the representation of the vector of BS to the RP $\bf{rp}$. The BS is set $[0,0]$ and the O is $[d_\text{UE-U}+d_\text{UE-O}\cos{\angle\text{O-UE-V}},d_\text{BS-U}-d_\text{UE-O}\sin{\angle\text{O-UE-V}}]$. Thus, the vector of $\bf{rp}$ equals to $[d_\text{UE-U}+d_\text{UE-O}\cos{\angle\text{O-UE-V}},d_\text{BS-U}-d_\text{UE-O}\sin{\angle\text{O-UE-V}}]$. Besides, the argument of $\bf{rp}$ can be deduced to $\angle\text{O-UE-V}-2\theta$ by utilizing the mirror point UE'. Consequently, the second equation can be written as:
\begin{equation}\label{sensing eq2}
\begin{split}
\arg{{\bf{rp}}}=\angle\text{O-UE-V}-2\theta.
\end{split}
\end{equation}

These two equations of Eq. \ref{sensing eq1} and Eq. \ref{sensing eq2} are utilized to solve the two values of $d_\text{UE-O}$ and $\theta$. $\theta$ equals the inclination of the tiny reflection face. Finally, the coordinates of RP O can be expressed, and the sketch of the environment would be reconstructed through the Os of all the MPCs at multiple UEs. The whole algorithm of ERM for all MPCs is outlined in Algorithm \ref{algorithm 1}. To reduce the environmental noise of the sensing, such as the wave through the passing pedestrians and the low grass, we identify the MPC with maximum power within each cluster. Then, a threshold $P_{thre_i}$ is calculated by subtracting two times the minimum reflection loss in the whole scenario from the free space path loss of the corresponding delay expressed as:
\begin{equation}\label{filt}
\begin{split}
P_{thre_i}=&PL_\text{FSPL}(D_{index_i}\times \text{c})-2\times \\
&\text{sort}(\min_i{PL_\text{FSPL}(D_{index_i}\times \text{c})-P_{index_i}})_a,
\end{split}
\end{equation}
\noindent where $PL_\text{FSPL}$ is introduced in \cite{sci_china}, $\text{c}$ is the speed of light, $\text{sort}(\cdot)$ represents sort the array in ascending order, and $a$ equals 2 and 1 in the line-of-sight (LoS) and non-LoS (NLoS) case, respectively, thereby facilitating filtration of available and one-order reflection MPCs from among all detected MPCs.
\begin{algorithm}
\caption{Solve the RPs of the available MPCs}\label{algorithm 1}
\begin{algorithmic}[1]
    \REQUIRE The number of MPCs $N$. The power, delay, and AoA of ALL MPCs for the one UEs, which are $P \ [\text{dB}]=\{P_\text{1},P_\text{2},P_\text{3},...,P_\text{N}\}$, $D \ [\text{s}]=\{D_\text{1},D_\text{2},D_\text{3},...,D_\text{N}\}$
    and $A \ [\text{rad}]=\{A_\text{1},A_\text{2},A_\text{3},...,A_\text{N}\}$, respectively. The coordinates of BS $[0,0]$ and UE $[a,b]$.
    \ENSURE The coordinates of RP O $[X_\text{0},Y_\text{0}]$.\\
    \STATE Clustering the MPCs by $P$, $D$, and $A$ into $M$ clusters.\\
    \STATE Find the index $index_\text{i}$, with $i$ from 1 to $M$, of MPC with the maximum power $P_{index_\text{i}}$ in the cluster $i$.
    \STATE Calculate the reflection loss $RL_{index_i}$ by ${PL_\text{FSPL}(D_{index_i}\times3\times10^{8})-P_{index_i}}$. 
    \STATE Get the threshold $P_{thre_i}$ by $PL_\text{FSPL}(D_{index_i}\times3\times10^{8})$ minus two times of the second minimum $RL_{index_i}$ in the LoS case and the minimum $RL_{index_i}$ in the NLoS case. 
    \FOR{$i=1; i \le M$}
    \IF{In the LoS case}
    \STATE Delete the LoS MPCs.
    \ENDIF
    \IF{$P_{index_\text{i}} > P_{thre}$}
    \STATE Solve the Eq. \ref{sensing eq1} and Eq. \ref{sensing eq2} and $D_{index_\text{i}}$ with $\angle\text{O-UE-V}=A_{index_\text{i}}$, $d_\text{BS-O-UE}=D_{index_\text{i}}\times3\times10^8$, $d_\text{B-U}=b$, and $d_\text{U-UE}=a$.  
    \STATE Get the $\theta$ and $d_\text{UE-O}$.
    \RETURN The coordinates of RP O for $i$ cluster $[a+d_\text{UE-O}\cos{A_{index_\text{i}}},b-d_\text{UE-O}\sin{A_{index_\text{i}}}]$.
    \ENDIF
    \ENDFOR
\end{algorithmic}
\end{algorithm}
\begin{figure}[!ht]
\centering
\subfigure[The layout of the simulation]{\includegraphics[width=4.5cm]{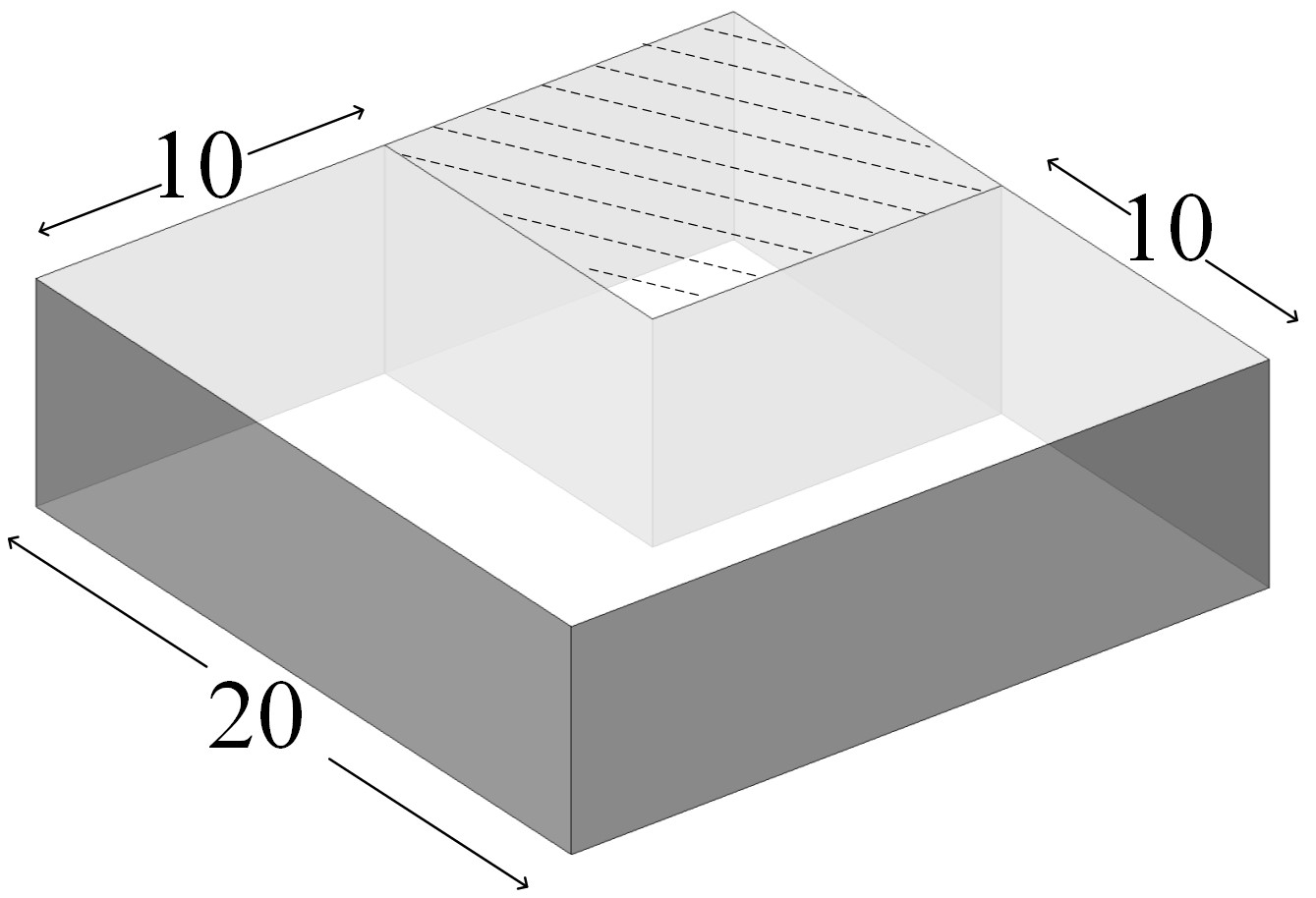}}
\subfigure[The sensing result]{\includegraphics[width=4.4cm, trim=7.4cm 4.1cm 2cm 4.3cm, clip]{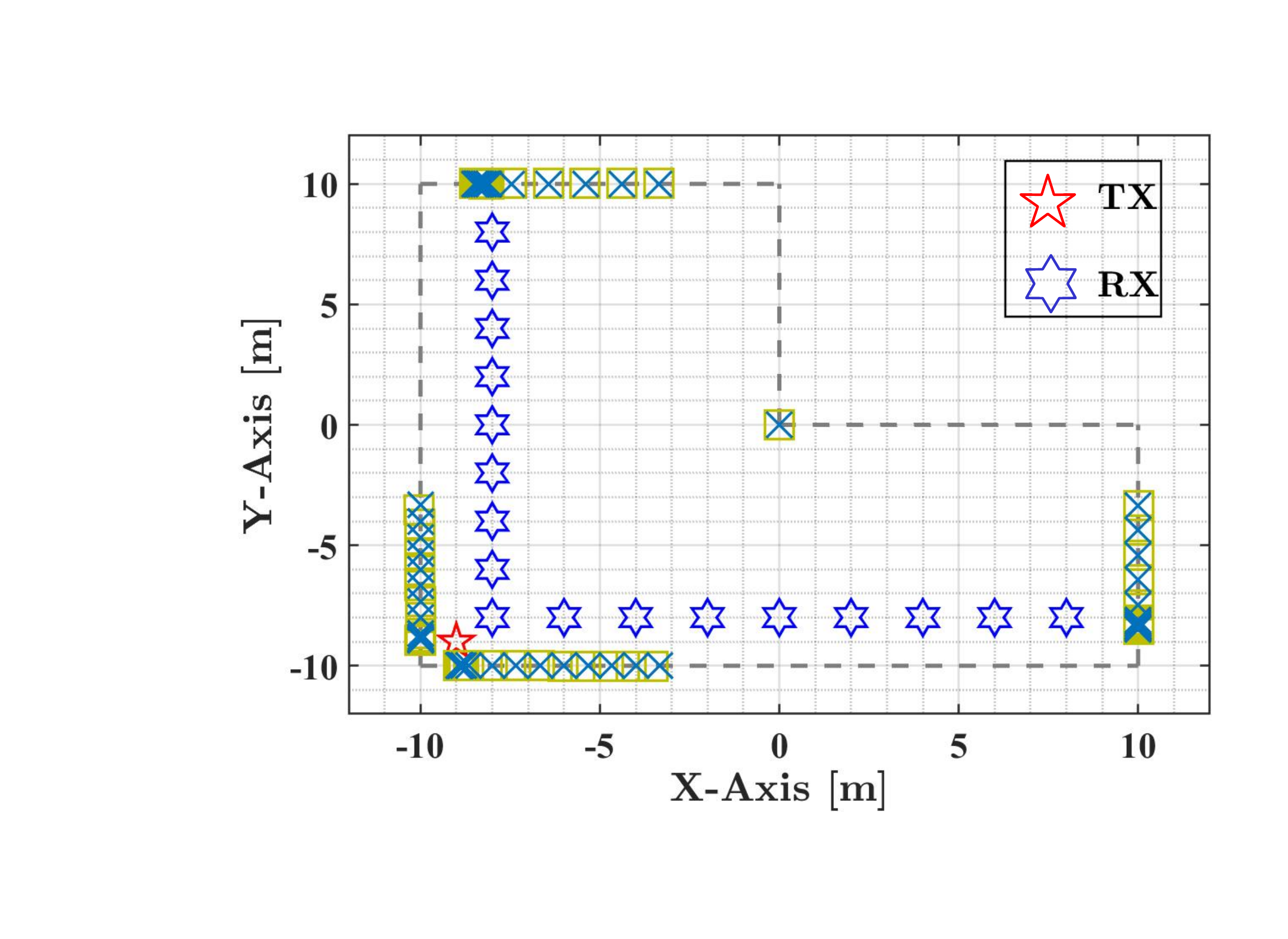}}
\subfigure[The CDF of reconstruction errors]{\includegraphics[width=4.25cm]{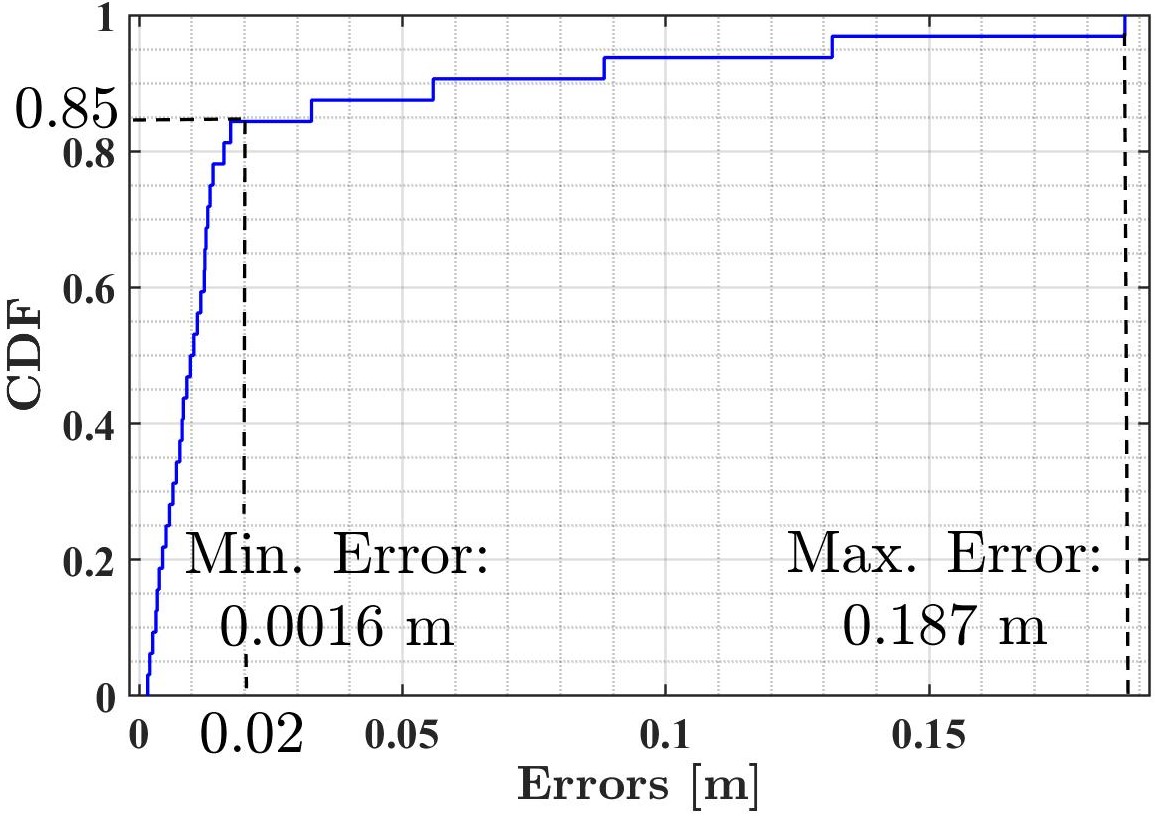}}
\caption{The layout (a), sensing result (b), and reconstruction errors CDF (c) of the ray-tracing simulation.}\label{simulation layout}
\end{figure}
\section{Verification With Ray-tracing Simulation}\label{2}
To validate the ERM, two experiments were conducted in THz bands: ray-tracing simulation in an L-shaped room and channel measurement in the UMa scenario. Those propagation environments were reconstructed based on the experiment results. In this section, the verification based on ray-tracing simulation is first introduced to demonstrate the fundamental correctness of the sensing algorithm. 
\begin{table}[t]
\begin{center}
\setlength{\tabcolsep}{1.8pt}  %%% ´Ë²ÎÊýÓÃÓÚµ÷Õû±íµÄÕûÌå¿í¶È
\renewcommand\arraystretch{1}  %% µ÷Õû±íÄÚÐÐÓëÐÐÖ®¼äµÄ×ÝÏò¾àÀë
\caption{THE SETUP OF RAY-TRACING SIMULATION AND CHANNEL MEASUREMENT.}\label{Table 1}
\vglue8pt
\begin{tabular}{ccc}  %% Ö¸¶¨Ò»¸öÓÐ6ÁÐÊý¾ÝµÄ±í£¬cÎªÁÐ¾ÓÖÐ, l ÎªÁÐ×ó¶ÔÆë, r ÎªÁÐÓÒ¶ÔÆë
 \hline
  {\textbf{Parameter}}   &{\textbf{Simulation Value}} &{\textbf{Measurement Value}}\\     %% µÚ1 ÐÐ
 \hline
   {TX/RX height}   &{4/2 $\rm{m}$}&{28.4/1.4 $\rm{m}$}\\
    \hline
   {Centre frequency}   &{300 $\rm{GHz}$}&{132 $\rm{GHz}$}\\
    \hline
     {Bandwidth}     &{1.2 $\rm{GHz}$}&{1.2 $\rm{GHz}$}\\
    \hline
   {TX/RX Antenna gain}    &{0 $\rm{dBi}$}&{24.5 $\rm{dBi}$}\\
    \hline
   {\makecell{Horizontal/vertical \\ HPBW at TX and RX}}    &{360$^\circ$/180$^\circ$}&{10$^\circ$/9$^\circ$}\\
    \hline
   {RX Azimuth rotation range}  &{-} &{[0$\degree$: 10: 360$\degree$)}\\
    \hline
   {RX Elevation rotation range} &{-} &{[-9$\degree$: 9: 9$\degree$]} \\
    \hline
\end{tabular}
\end{center}
\end{table}

The configuration of the ray-tracing simulation is shown in Table \ref{Table 1}. The chosen scenario consists of an L-shaped room with dimensions of 20$\times$10 m, as depicted in Fig. \ref{simulation layout}(a). The electromagnetic parameters of the plaster wall, for example, the correlation length, height standard deviation, and refractive index are 0.15, 1.7, and 1.92-j0.059, respectively, which are detailed in \cite{em_p}. Both the transmitter (TX) and RX utilize omnidirectional antennas with a gain of 0 $\rm{dBi}$. Fig. \ref{simulation layout}(b) shows the reflection points obtained from the ray-tracing simulation and our proposed sensing methodology, where the blue crosses denote simulating reflection points and the yellow squares represent sensing points. The reconstruction error is calculated by the Euclidean distance between the simulating reflection point and the sensing one. The cumulative distribution function (CDF) of reconstruction errors is shown in Fig. \ref{simulation layout}(c). The max and min errors are 0.187 and 0.0016 m, respectively. Almost 85$\%$ errors are smaller than 0.02 m. Moreover, the mean error and RMSE are 0.02 m and 0.03 m, respectively, indicating an excellent performance of the proposed methodology.
\section{Verification With Channel Measurement}\label{3}
\subsection{THz channel measurement}\label{31}
In order to verify the accuracy of the ERM for the actual THz communication system, the deviations between the sensing RPs and the measured environment outline are calculated based on the channel measurement in the UMa scenario at 132 $\rm{GHz}$, which are the unexplored scenarios and frequency points within the bands (130-134 GHz) allocated by the ITU for mobile use\cite{chang_132}. 

Here, the structure of the channel sounder is detailed in the \cite{chang_132}. The frequency multiplying factor of the frequency multiplier is 6. The IF and LO frequencies are 6 and 21 $\rm{GHz}$, respectively. Besides, 1022 IQ signal samples are obtained with a sample rate of 1.2 $\rm{GHz}$. The antenna gains at TX and RX are 24.5 $\rm{dBi}$. The HPBWs of the Tx and RX antenna are 10$\degree$ and 9$\degree$ for horizontal and elevation planes. Further information on the channel sounder configuration for the measurement campaigns is shown in Table \ref{Table 1}.
\begin{figure}[!t]
\centering
\subfigure[The layout in the UMa]{\includegraphics[width=4.02cm, trim=7cm 21.5cm 5cm 3cm, clip]{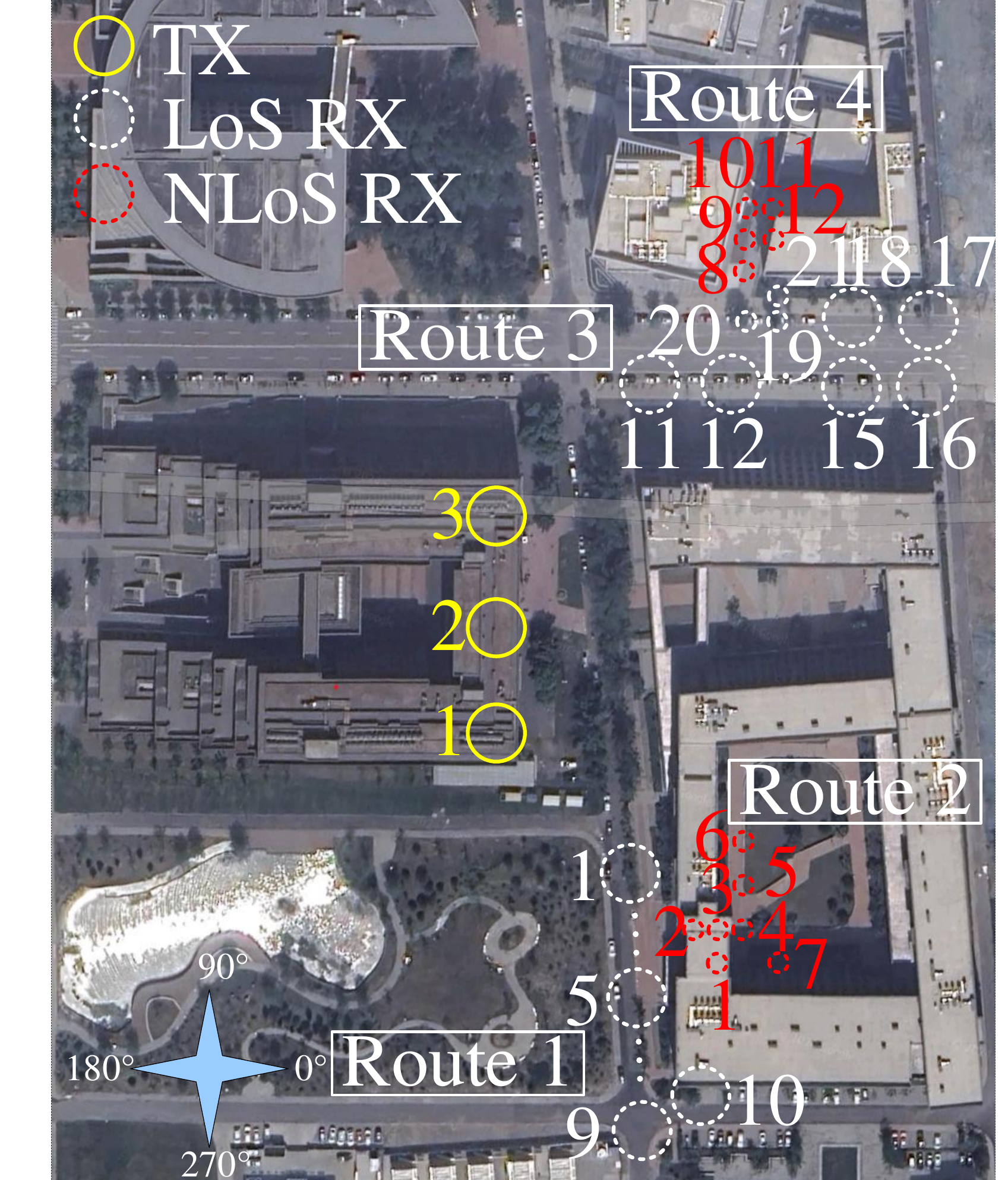}}
\subfigure[The measurement environment]{\includegraphics[width=3.92cm]{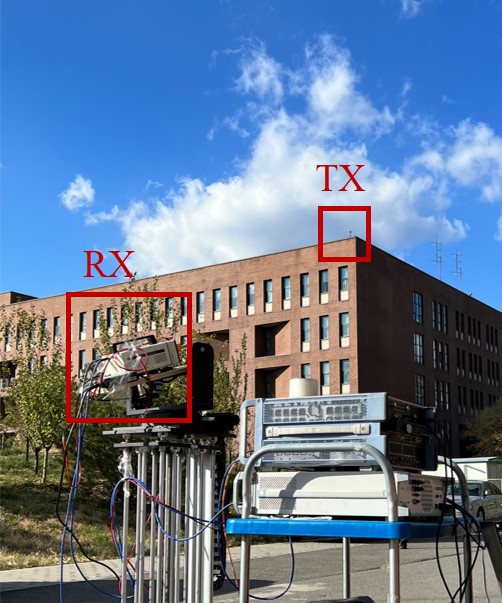}}
\caption{The measurement layout (a), and environment (b) in the UMa.}\label{measurement layout}
\end{figure}
\begin{figure}[!t]
\centering
\subfigure[$\rm{Route \ 1}$]{\includegraphics[width=4.02cm, trim=10.2cm 5.4cm 11cm 4cm, clip]{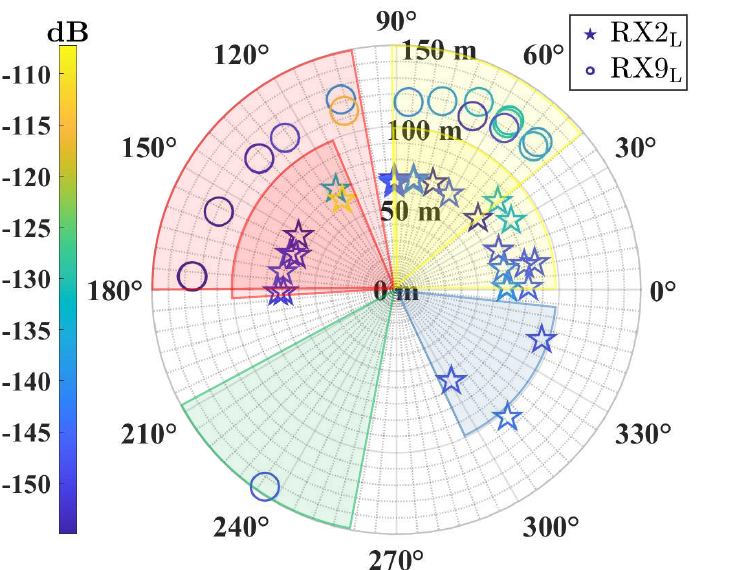}}
\subfigure[$\rm{Route \ 2}$]{\includegraphics[width=4.00cm, trim=10.2cm 5.4cm 11cm 4cm, clip]{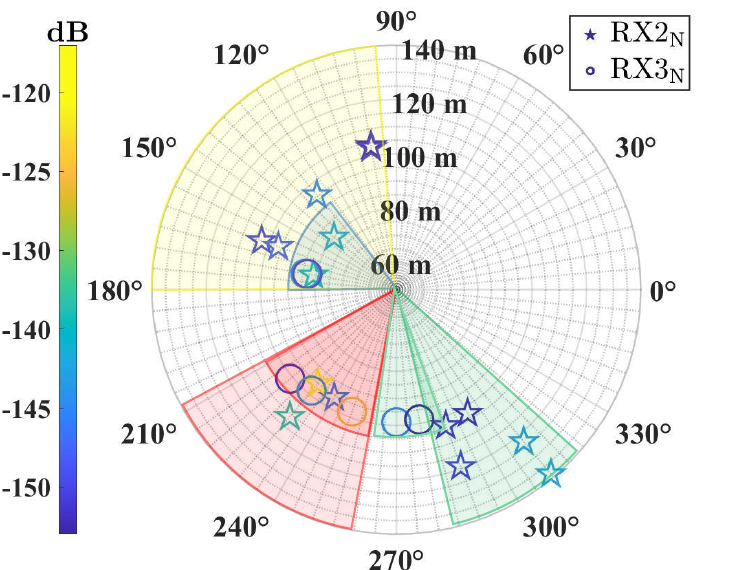}}
\caption{DAPSs of $\rm{RX2_L}$ and $\rm{RX9_L}$ in the route $1$ and $\rm{RX2_N}$ and $\rm{RX3_N}$ in the route $2$.}\label{DAPS}
\end{figure}
\begin{table}[t]
\begin{center}
\setlength{\tabcolsep}{6pt}  %%% ´Ë²ÎÊýÓÃÓÚµ÷Õû±íµÄÕûÌå¿í¶È
\renewcommand\arraystretch{1}  %% µ÷Õû±íÄÚÐÐÓëÐÐÖ®¼äµÄ×ÝÏò¾àÀë
\caption{THE DETAILS OF MEASUREMENT POINTS.}\label{Table 2}
\vglue8pt
\begin{tabular}{cccc}  %% Ö¸¶¨Ò»¸öÓÐ6ÁÐÊý¾ÝµÄ±í£¬cÎªÁÐ¾ÓÖÐ, l ÎªÁÐ×ó¶ÔÆë, r ÎªÁÐÓÒ¶ÔÆë
 \hline
  {\textbf{Route}}   &{\textbf{Tx}} &{\textbf{Rx}} &{\textbf{Distance [m]}}\\     %% µÚ1 ÐÐ
 \hline
   {Route 1}  &{$\rm{TX1}$} &{LoS 1-10} &{\makecell{62.2, 66.0, 70.0, 74.1, 78.3, 87.0,\\95.9, 105.0, 114.3, 111.9}}\\
    \hline
    {Route 2}  &{$\rm{TX1}$} &{NLoS 1-7} &{\makecell{83.4, 78.0, 80.6,\\83.3, 83.6, 83.1, 87.9}}\\
    \hline
     {Route 3}  &{$\rm{TX3}$} &{LoS 11-21} &{\makecell{57.3, 65.0, 73.3, 81.9, 90.8, 99.9,\\105.6, 97.0, 98.7, 92.3, 103.0}}\\
    \hline
      {Route 4}  &{$\rm{TX3}$} &{NLoS 8-12} &{\makecell{98.8, 102.0, 105.4, 108.2, 105.6}}\\
    \hline
\end{tabular}
\end{center}
\end{table}

The UMa scenario and locations of measurement points are illustrated in Fig. \ref{measurement layout}(a), while the measurement environment is shown in Fig. \ref{measurement layout}(b). TX is set at $\rm{TX1}$, $\rm{TX2}$, and $\rm{TX3}$ for both LoS and NLoS cases, with TX and RX heights consistently maintained at 28.4 and 1.4 $\rm{m}$, respectively, throughout the measurement. Additionally, a total of 23 RX locations ($\rm{RX1_L}$-$\rm{RX23_L}$) are set for the case of LoS condition and 12 RX locations ($\rm{RX1_{N}}$-$\rm{RX12_{N}}$) for the case of NLoS condition distributed across five routes as shown in Fig. \ref{measurement layout}(a). The distances between TX and RX are shown in Table \ref{Table 2}. 
\subsection{Mapping to the environment in spatial domain}\label{32}
This subsection explores the mapping of MPCs to the propagation environment, aiming at an initial investigation into the spatial domain correspondence correctness between them. Based on the channel measurement, the DAPS results for the representative routes are obtained \cite{chang_132} and shown in Fig. \ref{DAPS}, in which the MPCs are categorized into distinct groups based on the reflector conditions represented by colorful sectors. The color of the sample line represents the MPC power. For route 1, two DAPSs of $\rm{RX2_L}$ and $\rm{RX9_L}$ are combined in the one polar diagram of Fig. \ref{DAPS}a to present the correlation between the environment and ``angle clusters" of MPCs along with the route. Similarly, Fig. \ref{DAPS}b shows the DAPSs of $\rm{RX2_N}$ and $\rm{RX3_N}$ in the route 2. 

For route 1, except for the LoS red sector, it is observed that owing to the proximity of the teaching building along route 1, extensive MPCs at $0^\circ$-$90^\circ$ (yellow sector) and $270^\circ$-$360^\circ$ (blue sector) generate after the single reflection of the buildings at $\rm{RX2_L}$. Conversely, $\rm{RX9_L}$ does not detect MPCs within the range of $270^\circ$-$360^\circ$, as it lies beyond the southern boundary of the teaching building. Finally, it is worth noting that there exists potential for back-reflection from a closing wall at approximately $240^\circ$ within the green sector of $\rm{RX9_L}$. 

For route 2, the red sector covers the maximum power of both RXs and also includes MPCs reflected from the same reflector with maximum power. The green sectors are produced by the identical reflector points shortly after $\rm{RX3_N}$. The yellow sectors only appear in $\rm{RX2_N}$ but not in $\rm{RX3_N}$. The reason is that $\rm{RX3_N}$ can not receive the direction in the yellow sector in the case of being blocked by the closed north wall. Apart from the yellow and green sectors, the blue sector from $140^\circ$ to $170^\circ$ denotes a shorter delay than the other MPCs. This suggests that a near reflector exists between the TX and the reflector of the maximum power path, which would scatter the direct wave radiating from the TX. To sum up, the majority of MPCs in the THz channel could map the reflectors surrounding the RX in the spatial domain and do not undergo a diffraction process, which provides a significant gain in reconstructing the propagation environment. 
\subsection{Reconstruction results of channel measurement}\label{33}
\begin{figure}[!t]
\centering
\subfigure[Results of route 1 and 2]{\includegraphics[width=4cm, trim=25cm 22cm 35cm 12cm, clip]{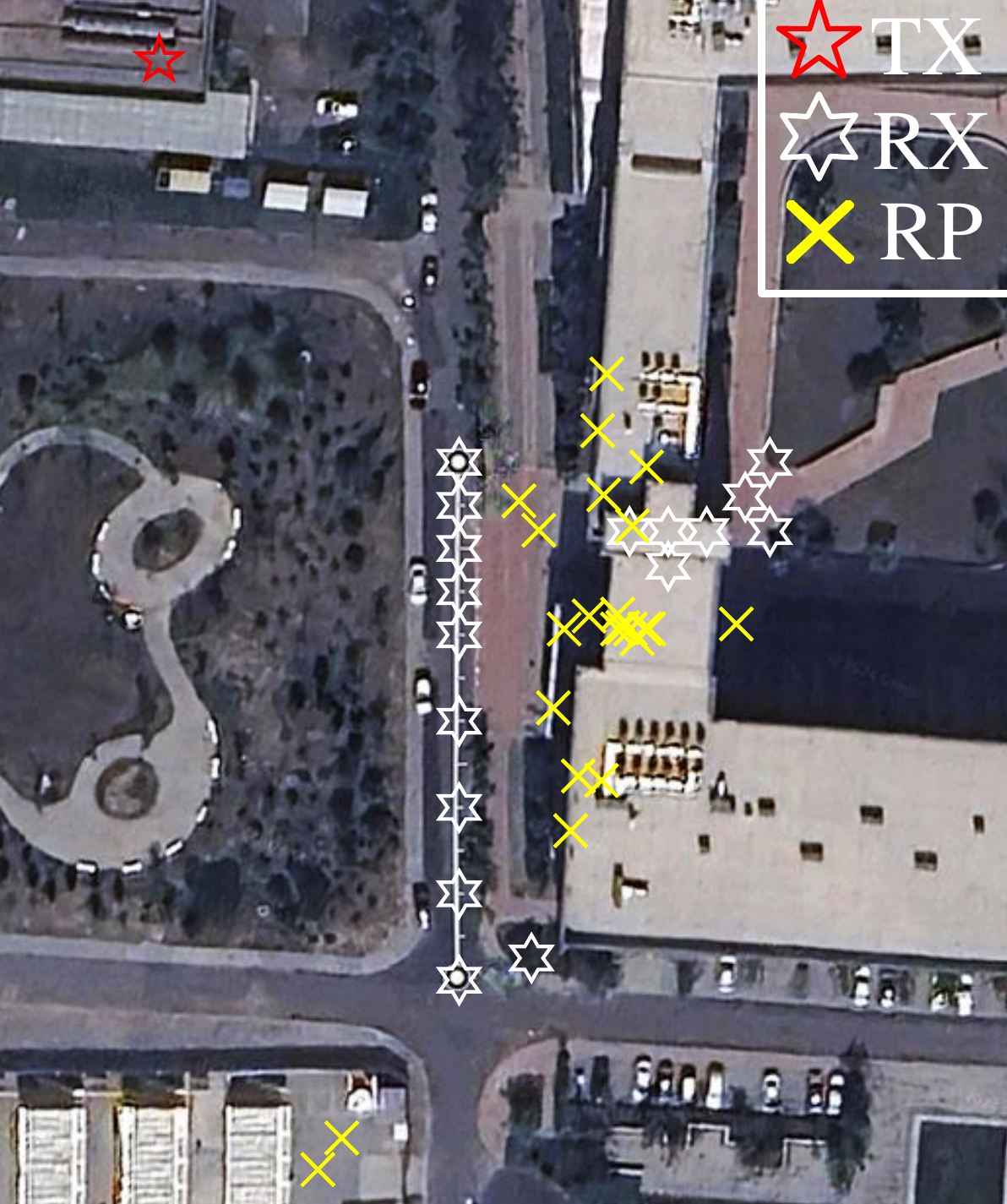}}
\subfigure[The sketch of error calculation]{\includegraphics[width=3.8cm, trim=3.5cm 2.4cm 18.2cm 1.5cm, clip]{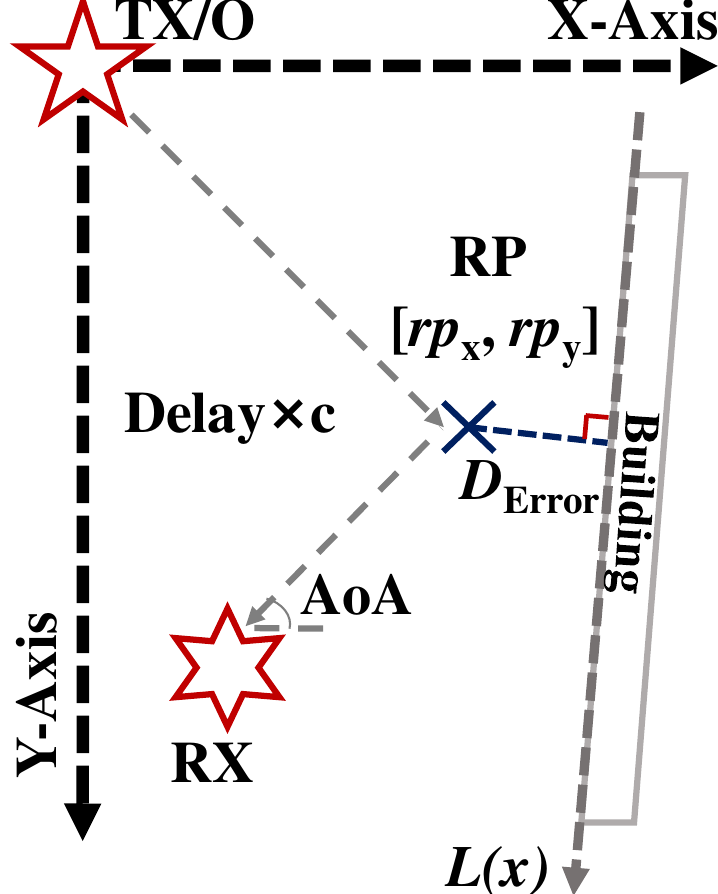}}
\caption{The results of routes 1, 2 (a) and the sketch of error calculation (b)}\label{sensing result}
\end{figure}

The sensing results of routes 1 and 2 are shown in Fig. \ref{sensing result}(a). The red star represents TX, the white hexagons are the RXs along routes 1 and 2, and the yellow crosses denote the sensed RPs. The Google Map is utilized to provide a reference with a uniform scale. It is evident that the building outline, including a wide corridor measuring approximately $14.9$ $\rm{m}$ under the roof, can be reconstructed. Additionally, two RPs above the street are identified, meaning that the material lamp standard is an indispensable reflector for NLoS RXs to receive enough power. The southwest residence also offers several opportunities for reflections toward nearby RXs which has been sensed. 

The deviations between the RPs of reconstruction results and the measured environment outline are calculated to evaluate the accuracy of the proposed sensing approach. The precision of the RPs is evaluated by computing the discrepancy of the shortest distances between the sensing outcomes and the measured environment outline as shown in Fig. \ref{sensing result}(b). The coordinate of TX is set to $[0,0]$, with the east and south directions assigned as the positive direction of the x-axis and y-axis, respectively, in routes 1 and 2. The building faces can be expressed as a linear equation $L$: 
\begin{equation}\label{error2}
\begin{split}
L(x)=a_\text{L}x+a_\text{L},
\end{split}
\end{equation}
\noindent where the $L$, $a_\text{L}$, and $b_\text{L}$ denote the y-axis, slope, and intercept of the building face, respectively, and the $x$ represents its x-axis. The reconstruction error is evaluated by the vertical distance between the RPs [$rp_\text{x}$,$rp_\text{y}$] and the building face, which is written as:
\begin{equation}\label{error3}
\begin{split}
D_\text{Error}=\frac{|rp_\text{y}+a_\text{L}\times rp_\text{x}-b_\text{a}|}{1+a_\text{L}^2}.
\end{split}
\end{equation}

For instance, the coordinates of measured points at the east building (EB) are $[46.78,46.57]$ and $[41.1,106.57]$. The vector of the EB surface $[-5.68,60.00]$ can be obtained by these points, and $a_\text{L}$ and $a_\text{L}$ are $-10.56$ and $540.72$, respectively. Besides, the sensing RPs coordinates $[rp_x,rp_y]$ at the EB are $[46.4,36.35]$, $[45.45,42.91]$, $[46.09,50.36]$, $[50.53,47.11]$, $[44.58,64.38]$, $[47.51,65.90]$, $[40.87,75.03]$, $[43.46,82.92]$, $[45.87,83.43]$, and $[42.66,89.26]$. Thus, the reconstruction errors can be computed and summarized in Table \ref{Table 3}. The RMSE of the reconstruction errors is $1.28$ $\rm{m}$. The reason for the error is that the surface of the EB contains windows and is not an ideal smooth plane. In the same way, the error of reconstructing the corridor wall in the NLoS case can be calculated. The walls in this case are horizontal and the invariable ordinate of double walls (shorter sidewalk wall and teaching building wall) of the north corridor wall and one wall of the south corridor wall are $53.38$, $50.38$, and $65.30$, respectively. The ordinate of sensing points are $53.94$, $65.80$, $65.73$, $65.89$, $65.90$, $65.86$, $65.31$, and $67.01$. It is worth mentioning that only one MPC reflected from the sidewalk wall in the north direction can be detected. The RMSE of these sensing is $0.45$ $\rm{m}$. The smooth walls guarantee better sensing results than the EB one.

Besides, in the same way as routes 1 and 2, the deviations of the ERM of routes 3 and 4 are also listed in Table \ref{Table 3}, including the porch located before the north building at the upper of route 3 and the street ``canyon'' in route 4. The RMSE of sensing deviation for the walls of the ``canyon" and porch are 1.28 and 0.51 $\rm{m}$, respectively. The big deviation in sensing the canyon walls indicates their uneven plane constructed by doors, windows, and green plants. To sum up, the errors of route 2 and porch are the smallest and the ``canyon" presents the largest maximum, minimum, and mean errors due to those RXs with similar reflecting points.  
\begin{table}[t]
\begin{center}
\setlength{\tabcolsep}{7pt}  %%% ´Ë²ÎÊýÓÃÓÚµ÷Õû±íµÄÕûÌå¿í¶È
\renewcommand\arraystretch{1}  %% µ÷Õû±íÄÚÐÐÓëÐÐÖ®¼äµÄ×ÝÏò¾àÀë
\caption{THE SUMMARIZATION OF SENSING ERROR.}\label{Table 3}
\vglue8pt
\begin{tabular}{ccccc}  %% Ö¸¶¨Ò»¸öÓÐ6ÁÐÊý¾ÝµÄ±í£¬cÎªÁÐ¾ÓÖÐ, l ÎªÁÐ×ó¶ÔÆë, r ÎªÁÐÓÒ¶ÔÆë
 \hline
  {\textbf{Scenario}}   &{\textbf{Route 1}} &{\textbf{Route 2}} &{\textbf{Canyon}} &{\textbf{Porch}}\\     %% µÚ1 ÐÐ
 \hline
     {\textbf{Min. $D_\text{Error}$ [m]}}  &{0.09} &{0.01} &{1.00} &{0.37}\\
    \hline
    {\textbf{Max. $D_\text{Error}$ [m]}}  &{3.76} &{1.71} &{4.44} &{1.87}\\
    \hline
   {\textbf{Mean. $D_\text{Error}$ [m]}}  &{1.62} &{0.62} &{2.52} &{1.10}\\
    \hline
     {\textbf{RMSE [m]}}  &{1.28} &{0.45} &{1.28} &{0.51}\\
    \hline
\end{tabular}
\end{center}
\end{table}
\section{CONCLUSION}\label{5}
This paper focuses on proposing the ERM for reconstructing the propagation environment using small-scale parameters of the receiving radio and validating it through ray-tracing simulation and channel measurement in the UMa scenario in THz bands. The verification of the ray-tracing simulation yields an RMSE of 0.03 m. For measured verification, we investigate the mapping relation of MPCs to the realistic environment in the space domains by DAPS. It is observed that the RX side by a longer building presents a wider receiving angle domain than the RX close to the crossroad. Furthermore, small sensing RMSEs of 1.28 and 0.45 $\rm{m}$ in routes 1 and 2, and 1.28 and 0.51 $\rm{m}$ for walls of the ``canyon" and porch in routes 3 and 4 indicate a high accuracy of our method. Overall, this work provides valuable insight into ISAC in the THz bands. In future work, denser measurement points and rotation steps are needed to produce a more comprehensive outline of the environment.
\bibliographystyle{IEEEtran}
\bibliography{bib}
\end{document}